\documentclass[a4paper]{jpconf}
\usepackage{graphicx}
\begin{document}
\title{Strong-disorder magnetic quantum phase transitions: Status and new developments}

\author{T Vojta}

\address{Department of Physics, Missouri University of Science $\&$ Technology, Rolla, MO 65409, USA}

\ead{vojtat@mst.edu}

\begin{abstract}
This article reviews the unconventional effects of random disorder on magnetic quantum phase transitions,
focusing on a number of new experimental and theoretical developments during the last three years. On the theory side,
we address smeared quantum phase transitions tuned by changing the chemical composition,
for example in alloys of the type A$_{1-x}$B$_x$. We also discuss how the interplay of order parameter
conservation and overdamped dynamics leads to enhanced quantum Griffiths singularities in disordered metallic
ferromagnets. Finally, we discuss a semiclassical theory of transport properties in quantum Griffiths phases.
Experimental examples include the ruthenates Sr$_{1-x}$Ca$_x$RuO$_3$ and (Sr$_{1-x}$Ca$_x$)$_3$Ru$_2$O$_7$
as well as Ba(Fe$_{1-x}$Mn$_x$)$_2$As$_2$.
\end{abstract}

%%%%%%%%%%%%%%%%%%%%%%%%%%%%%%%%%%%%%%%%%%%%%%%%%%%%%%%%%%%%%%%%%%%%%%%%%%%%%%%%
\section{Introduction}
\label{sec:intro}
%%%%%%%%%%%%%%%%%%%%%%%%%%%%%%%%%%%%%%%%%%%%%%%%%%%%%%%%%%%%%%%%%%%%%%%%%%%%%%%%

The question of how random disorder influences a zero-temperature quantum phase transition has been
a subject of great interest during the last two decades. Pioneering theoretical work initially
focused on model systems such as the random transverse-field Ising chain. Using a strong-disorder renormalization
group, Fisher \cite{Fisher92,Fisher95} showed that the ferromagnetic quantum phase transition in this
system is of exotic infinite-randomness type (for a review of this method, see, e.g., Ref.\ \cite{IgloiMonthus05}).
It is accompanied by strong off-critical singularities in various thermodynamic quantities, the so-called
quantum Griffiths singularities. They are caused by rare spatial regions that are locally in the ``wrong phase.''
Analogous results were obtained using optimal fluctuation theory and
numerical simulations \cite{ThillHuse95,RiegerYoung96,YoungRieger96}. Later, these findings were generalized
to higher dimensions \cite{SenthilSachdev96,PYRK98,MMHF00}.

Castro Neto and Jones \cite{CastroNetoJones00} pioneered the application of the Griffiths singularity concept
to itinerant (metallic) systems. The qualitative behavior of disordered metals close to a magnetic instability
is determined by the order parameter symmetry together with the dissipative character of the order parameter
dynamics \cite{Vojta06}. For Ising symmetry, the quantum dynamics of a sufficiently large individual magnetic
rare region stops completely because the magnetic fluctuations are overdamped
\cite{MillisMorrSchmalian01}. This destroys the sharp global quantum phase transition by smearing \cite{Vojta03a,HoyosVojta08}.
In contrast, for continuous (XY or Heisenberg) order parameter symmetry, individual rare regions continue to
fluctuate to the lowest temperatures \cite{VojtaSchmalian05}; this leads to quantum Griffiths singularities similar to those of the
dissipationless transverse-field Ising chain.\footnote{This means, somewhat surprisingly, the dissipationless random transverse-field Ising
model and the disordered metallic Heisenberg magnet are in the same universality class.} Note that long-range interactions between different rare
regions can actually lead to a (cluster-glass) freezing transition at the lowest temperatures even in the
continuous symmetry case \cite{DobrosavljevicMiranda05}. These and many more results were summarized in Refs.\
\cite{Vojta06,Vojta10}.

While significant progress was made since the mid-1990s in the theory of disordered quantum phase transitions,
experimental verifications of the predicted exotic disorder effects were lacking for a long time. This changed
only in the years 2007 to 2010, when quantum Griffiths singularities were observed in the vicinity of ferromagnetic
quantum phase transitions in the magnetic semiconductor Fe$_{1-x}$Co$_x$S$_2$
\cite{GYMBHCHD07,GYMBHCHD10a,GYMBHCHD10b},
the Kondo lattice system  CePd$_{1-x}$Rh$_x$ \cite{GYMBHCHD10a,Westerkampetal09}, and the transition
metal alloy Ni$_{1-x}$V$_x$ \cite{UbaidKassisVojtaSchroeder10,SchroederUbaidKassisVojta11}.
These experiments were reviewed in Ref.\ \cite{Vojta10}.

In the last three years, the field has made further progress, both in theory and in experiment. The purpose
of the present paper is to review these new results and to relate them to the state of knowledge. The
paper is organized as follows. Section \ref{sec:theory} is devoted to new developments in the theory of
disordered quantum phase transitions. In particular, we address smeared quantum phase transitions
tuned by changing the chemical composition in systems of the type A$_{1-x}$B$_x$. We also consider the enhancement
of ferromagnetic quantum Griffiths singularities in disordered metallic systems as well as
a semiclassical theory of transport properties in quantum Griffiths phases.
In Sec.\ \ref{sec:experiment}, we review a number of exciting new experiments that show
additional strong disorder effects. In particular, we discuss the smeared quantum phase
transition in the ruthenate Sr$_{1-x}$Ca$_x$RuO$_3$, an unusual nearly ferromagnetic phase in (Sr$_{1-x}$Ca$_x$)$_3$Ru$_2$O$_7$,
as well as a possible antiferromagnetic quantum Griffiths phase in Ba(Fe$_{1-x}$Mn$_x$)$_2$As$_2$.
We conclude in Sec.\ \ref{sec:conclusions}.

%%%%%%%%%%%%%%%%%%%%%%%%%%%%%%%%%%%%%%%%%%%%%%%%%%%%%%%%%%%%%%%%%%%%%%%%%%%%%%%%
\section{New developments: Theory}
\label{sec:theory}
%%%%%%%%%%%%%%%%%%%%%%%%%%%%%%%%%%%%%%%%%%%%%%%%%%%%%%%%%%%%%%%%%%%%%%%%%%%%%%%%

\subsection{Composition-tuned smeared quantum phase transitions}
\label{subsec:compo}

Let us start by contrasting sharp and smeared transitions. Phase transitions are commonly defined as
singularities of the free energy in parameter space. Such singularities can only occur in infinitely
large systems; for continuous transitions they are associated with an infinite correlation length;
and in the case of first-order transitions, they correspond to macroscopic phase coexistence.

What about phase transitions in the presence of randomness that locally favors one phase over the
other? Do they remain sharp in the above sense, or will the singularity be smeared over a range of the
tuning parameter? It turns out that phase transitions in random systems generically remain sharp.
The reason is that finite-size regions normally cannot order independently
of each other; the global phase transition thus occurs due to a collective effect of the entire
sample. It was shown in Ref.\ \cite{Vojta03a}, however, that this argument does not apply to certain
magnetic quantum phase transitions in the presence of both disorder and dissipation. If a magnetic (Ising) cluster
is coupled to an Ohmic dissipative bath, it can undergo the localization transition
of the dissipative two-level system \cite{LCDFGZ87,Weiss_book93}. This implies that
sufficiently large clusters develop static magnetic order independently of the bulk because their quantum dynamics
stops completely. As more and more clusters freeze, the global order parameter thus arises gradually,
rather than as a collective effect of the entire system; and the correlation length remains finite. This defines
a smeared phase transition. The prime experimental application of this
concept are metallic quantum magnets in which the dissipation stems from the coupling of the
magnetization to the fermionic degrees of freedom.

It is important to emphasize that the smearing of the quantum phase transition is a finite-length-scale
phenomenon (there is a \emph{finite} minimum size required for the magnetic clusters to freeze),
in contrast to the usual critical phenomena that arise in the limit of infinite lengths. This also
means that smeared transitions cannot be expected to display the same degree of universality as
critical points. In particular, tuning the transition with different tuning parameters can lead to
qualitatively different behavior.

\begin{figure}
\centerline{\includegraphics[width=8cm]{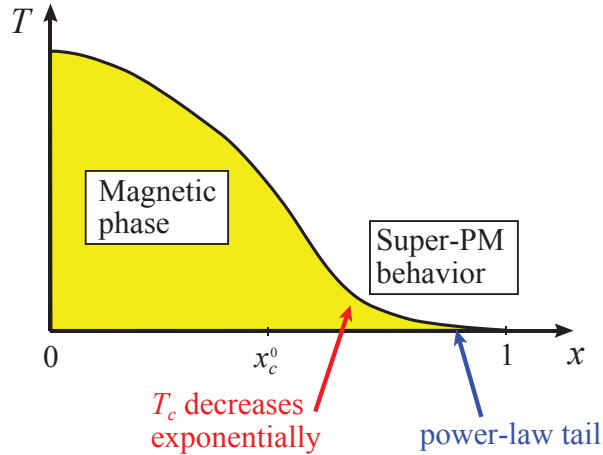}}
\caption{Schematic temperature-composition phase diagram of a binary alloy A$_{1-x}$B$_x$
    displaying a smeared quantum phase transition. In the tail of the magnetic phase,
    which stretches to $x=1$, the rare regions are aligned. Above $T_c$,
    they act as independent classical moments, resulting in super-paramagnetic (PM)
    behavior (after Ref.\ \cite{HrahshehNozadzeVojta11}).}
\label{fig:composition_pd}
\end{figure}

The original theory of smeared (magnetic) quantum phase transitions \cite{Vojta03a}
focused on tuning a given disordered sample through the transition by varying a parameter such as
pressure. However, in almost all of the experimental systems in which unconventional disorder effects
have been observed, the transition is actually tuned by changing the chemical composition. In Ref.\
\cite{HrahshehNozadzeVojta11}, we therefore adapted the theory to alloys of the type A$_{1-x}$B$_x$
which are tuned through a quantum phase transition by changing $x$. A schematic of the resulting
temperature-composition phase diagram is shown in Fig.\ \ref{fig:composition_pd}.
The ordered phase develops a pronounced tail that stretches all the way to $x=1$. Two separate
regimes can be distinguished. The major decrease of the critical temperature $T_c$ near the mean-field
critical concentration $x_c^0$ is of exponential type,
\begin{equation}
T_c \sim \exp\left [ -C \frac {(x-x_c^0)^{2-d/\phi}}{x(1-x)}\right ]
\label{eq:Tc-exponential}
\end{equation}
where $C$ is a nonuniversal constant and $\phi$ is the finite-size shift exponent.
In the far tail of the smeared transition, $x\to 1$, the critical temperature vanishes as a
nonuniversal power law,
\begin{equation}
T_c \sim (1-x)^{L_{\rm min}^d} = (1-x)^{(D/x_c^0)^{d/\phi}}
\label{eq:Tc-power}
\end{equation}
where $L_{\rm min}$ is the minimum size a rare region must have to freeze, and $D$ is another constant.
The zero-temperature magnetization $M$ behaves in a similar fashion. Close to $x_c^0$, $M$ varies
as
\begin{equation}
M \sim \exp\left [ -C \frac {(x-x_c^0)^{2-d/\phi}}{x(1-x)}\right ]
\label{eq:M-exponential}
\end{equation}
while it follows the power law
\begin{equation}
M \sim (1-x)^{L_{\rm min}^d} = (1-x)^{(D/x_c^0)^{d/\phi}}
\label{eq:M-power}
\end{equation}
for $x\to 1$.
We thus find that $M$  is nonzero in the entire composition range $0\le x<1$, illustrating the
notion of a smeared quantum phase transition.

The fact that smeared phase transitions are nonuniversal also implies that spatial disorder
correlations can have significant effects. This question was explored in Ref.\ \cite{SNHV12}. It was found
that even short-range correlations can qualitatively change the behavior of observable quantities
at smeared transitions, in contrast to critical points, at which short-range correlations do not
change the critical behavior.

\subsection{Quantum Griffiths singularities in ferromagnetic metals}

The theory of Griffiths singularities started when Griffiths \cite{Griffiths69} showed that the free energy
of a disordered systems that does undergo a sharp phase transition is singular not only right at the
transition point but in an entire parameter
region around out. Today, this region is known as the Griffiths region or Griffiths phase, and the off-critical
singularities are summarily called the Griffiths singularities.

The theory of quantum Griffiths phenomena in metallic systems initially focused on
antiferromagnetic quantum phase transitions. This had two reasons: First, there are many more materials
featuring antiferromagnetic than ferromagnetic quantum phase transitions. Second, the theory for
the antiferromagnetic case is considerably simpler (at least in three space dimensions). In the
ferromagnetic case, extra complications arise from the order parameter conservation and from
soft-mode induced long-range interaction between the order-parameter fluctuations
\cite{KirkpatrickBelitz96b,BelitzKirkpatrickVojta05}. Ironically, all the recent experimental observations
of quantum Griffiths physics mentioned in Sec.\ \ref{sec:intro} were in ferromagnetic
conductors (Fe$_{1-x}$Co$_x$S$_2$,
CePd$_{1-x}$Rh$_x$, Ni$_{1-x}$V$_x$) for which the theory was not well developed.

To remedy this problem, a theory of quantum Griffiths singularities associated with
itinerant ferromagnetic quantum phase transitions was developed in Ref.\ \cite{NozadzeVojta12}.
The crucial technical difference between the ferromagnetic and antiferromagnetic cases lies in the dynamic part of the
two-point vertex of the order-parameter (LGW) field theory \cite{Hertz76}. In the antiferromagnetic case,
it takes an Ohmic form $\sim |\omega_n|$ and becomes momentum-independent in the limit
of small momenta and frequencies. In a disordered ferromagnet, the damping term
instead has the structure $\sim |\omega_n|/\mathbf{q}^2$, reflecting the conserved nature
of the total magnetization. As a result, the tunneling rate of a locally ordered
ferromagnetic cluster decays as $\exp(-c L^5)$ with its linear size $L$, compared with
a $\exp(-c L^3)$ decay in the antiferromagnetic case ($c$ is a constant).

The form of the ferromagnetic quantum Griffiths singularities can now be worked out by
adapting the usual optimal fluctuation arguments (see, e.g., Ref.\ \cite{Vojta06}) to the
case at hand. The resulting low-energy density of states of the rare regions takes the form
\begin{equation}
\rho(\epsilon)\sim\frac{1}{\epsilon} \exp[\{ - \tilde{\lambda} \log(\epsilon_0/\epsilon)\}^{3/5}]\,.
\label{eq:DOS}
\end{equation}
Here, $\epsilon_0$ is a microscopic energy scale, and the nonuniversal parameter
$\tilde{\lambda}$ takes the place of the usual Griffiths exponent. In contrast to itinerant antiferromagnets
(and models such as the transverse-field Ising chain), this quantum Griffiths singularity in the
density of states is not a pure power law. However, the $\exp[|\ln \epsilon|^{3/5}]$ structure
implies that it is ``almost a power law.'' Correspondingly, the quantum Griffiths singularities
in observables also differ from pure power laws. The low-temperature susceptibility, for example,
is expected to behave as
\begin{equation}
 \chi(T) \sim \frac{1}{T} \exp[\{-\tilde{\lambda} \log(T_0/T)\}^{3/5}]\,.
\label{eq:susceptibility}
\end{equation}
Compared to the power-law form $T^{\lambda-1}$, eq.\ (\ref{eq:susceptibility}) provides an additional
upturn at low temperatures. This may be responsible for the slight upward deviations from power-law behavior
observed at the lowest temperatures in the susceptibility data of
Ni$_{1-x}$V$_x$ \cite{UbaidKassisVojtaSchroeder10,SchroederUbaidKassisVojta11}.
We emphasize that eqs.\ (\ref{eq:DOS}) and (\ref{eq:susceptibility}) only hold in the asymptotic regime, when the
relevant rare regions are so large that the order-parameter conservation significantly hampers their
dynamics.

\subsection{Semiclassical transport theory}

Most theoretical studies of disordered quantum phase transitions and the accompanying quantum Griffiths
singularities have focused on thermodynamics while transport properties have received much less attention.
This is particularly true for metallic systems whose transport properties such as the electrical resistivity
are easily accessible in experiments but hard to calculate theoretically.
In Ref.\ \cite{NozadzeVojta11}, we therefore developed a theory of the electrical resistivity in an antiferromagnetic
quantum Griffiths phase of a disordered metal. It is based on a semiclassical approach in which the conduction
electrons are scattered by the spin fluctuations of the rare regions. Other transport quantities including the
thermal resistivity, the thermopower and the Peltier coefficient were calculated in the same way.

The calculation starts from a two-band model of s and d electrons, $H=H_s + H_d +H_{s-d}$ \cite{UedaMoriya75}.
Only the s electrons contribute to the electric transport; they are scattered by the spin fluctuations of the
d electrons which are assumed to be in an antiferromagnetic quantum Griffiths phase. This is described by means of a Boltzmann
equation in which the dynamic susceptibility of the d electrons enters the collision term for the s electrons. The linearized
Boltzmann equation is then solved using Ziman's variational principle \cite{Ziman_book60}.
The resulting rare region contribution to the electrical resistivity takes the form
\begin{equation}
\Delta \rho \sim  T^\lambda \,.
\label{eq:rho}
\end{equation}
Thus, the temperature-dependence of the resistivity follows a non-universal power law controlled by the same
Griffiths exponent $\lambda$ that also governs the thermodynamics. (The reason is that the main
contribution to the resistivity stems from scattering off the magnetic fluctuations whose temperature
dependence is characterized by $\lambda$.) Analogous calculations for the thermal
resistivity $W$, the thermopower $S$, and the Peltier coefficient $\Pi$ yield
\begin{equation}
\Delta W \sim  T^{\lambda-1}, \quad \Delta S \sim  T^{\lambda+1}, \quad \Delta \Pi \sim  T^{\lambda+2} \,.
\label{eq:WSP}
\end{equation}
All of these results apply to antiferromagnets in three space dimensions for which quasiparticles are
(marginally) well-defined. Moreover, they hold in the case of strong randomness when the temperature-independent
impurity part $\rho_0$ of the resistivity is large compared to the rare region contribution (see Ref.\ \cite{Rosch99}).

The unusual temperature dependencies of thermodynamic and transport properties also affect various ``universal ratios''
that are commonly used to characterize Fermi liquid behavior \cite{NozadzeVojta12b}. The Wilson ratio $R_W = \chi T/C$
(which relates the susceptibility $\chi$ and the specific heat $C$) and the Gr\"uneisen parameter $\Gamma = \beta/C$
(the ratio of the thermal expansion coefficient $\beta$ and $C$) both diverge logarithmically with $T$,
\begin{equation}
R_W = \chi T/C \sim [\log(1/T)]^2~, \qquad \Gamma = \beta/C \sim \log(1/T)~.
\end{equation}
While $R_W$ and $\Gamma$ only involve thermodynamics, further ratios can be calculated from the transport quantities.
For example, the rare regions give a subleading contribution to the Lorenz number $L=\rho/WT$,
\begin{equation}
\Delta L \sim T^\lambda~.
\end{equation}
Moreover, the ratio $q=S/C$ between the thermopower and the specific heat as well as Kadowaki-Woods ratio
$R_{KW}=(\rho-\rho_{\rm imp})/C^2$ both show nontrivial power laws in $T$,
\begin{equation}
q =S/C \sim T^{1-\lambda} ~, \qquad R_{KW} = (\rho-\rho_{\rm imp})/C^2 \sim T^{-\lambda}~,
\end{equation}
in contrast to the temperature-independent Fermi-liquid behavior.

The above results apply to quantum phase transitions in antiferromagnetic metals, the corresponding
theory for the more complicated ferromagnetic quantum phase transition remains a task for the future.

%%%%%%%%%%%%%%%%%%%%%%%%%%%%%%%%%%%%%%%%%%%%%%%%%%%%%%%%%%%%%%%%%%%%%%%%%%%%%%%%
\section{New developments: Experiment}
\label{sec:experiment}
%%%%%%%%%%%%%%%%%%%%%%%%%%%%%%%%%%%%%%%%%%%%%%%%%%%%%%%%%%%%%%%%%%%%%%%%%%%%%%%%

\subsection{Smeared quantum phase transition in Sr$_{1-x}$Ca$_x$RuO$_3$}

The perovskite-type ruthenate SrRuO$_3$ is a ferromagnetic metal having a Curie temperature $T_c^0$
of about 165\,K. In contrast, CaRuO$_3$ does not develop long-range magnetic order even at the lowest
temperatures. By randomly substituting the smaller calcium atoms for the strontium atoms, one can thus
tune the material through a ferromagnetic quantum phase transition. This system was extensively studied in
the past, but different experiments
\cite{CMSCG97,KYKMG99,KOCM04,Okamotoetal07,Hosakaetal08,SchneiderMoshnyagaGegenwart10,Wissingeretal11}
showed large variations of the phase boundary and especially in the critical calcium concentration $x_c$
beyond which long-range magnetic order vanishes.

\begin{figure}
\centerline{\includegraphics[width=8.8cm]{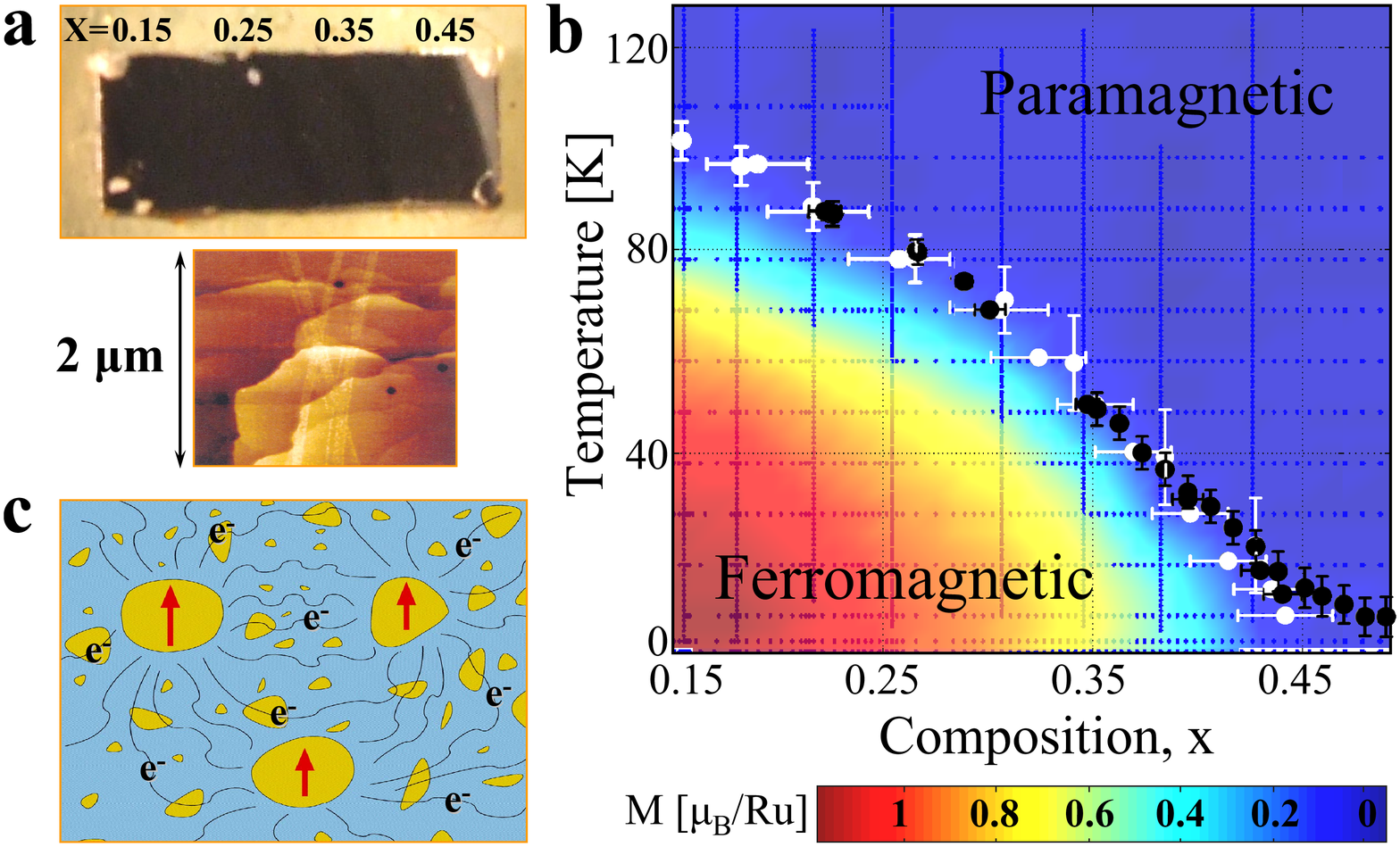}~~~\includegraphics[width=6.7cm]{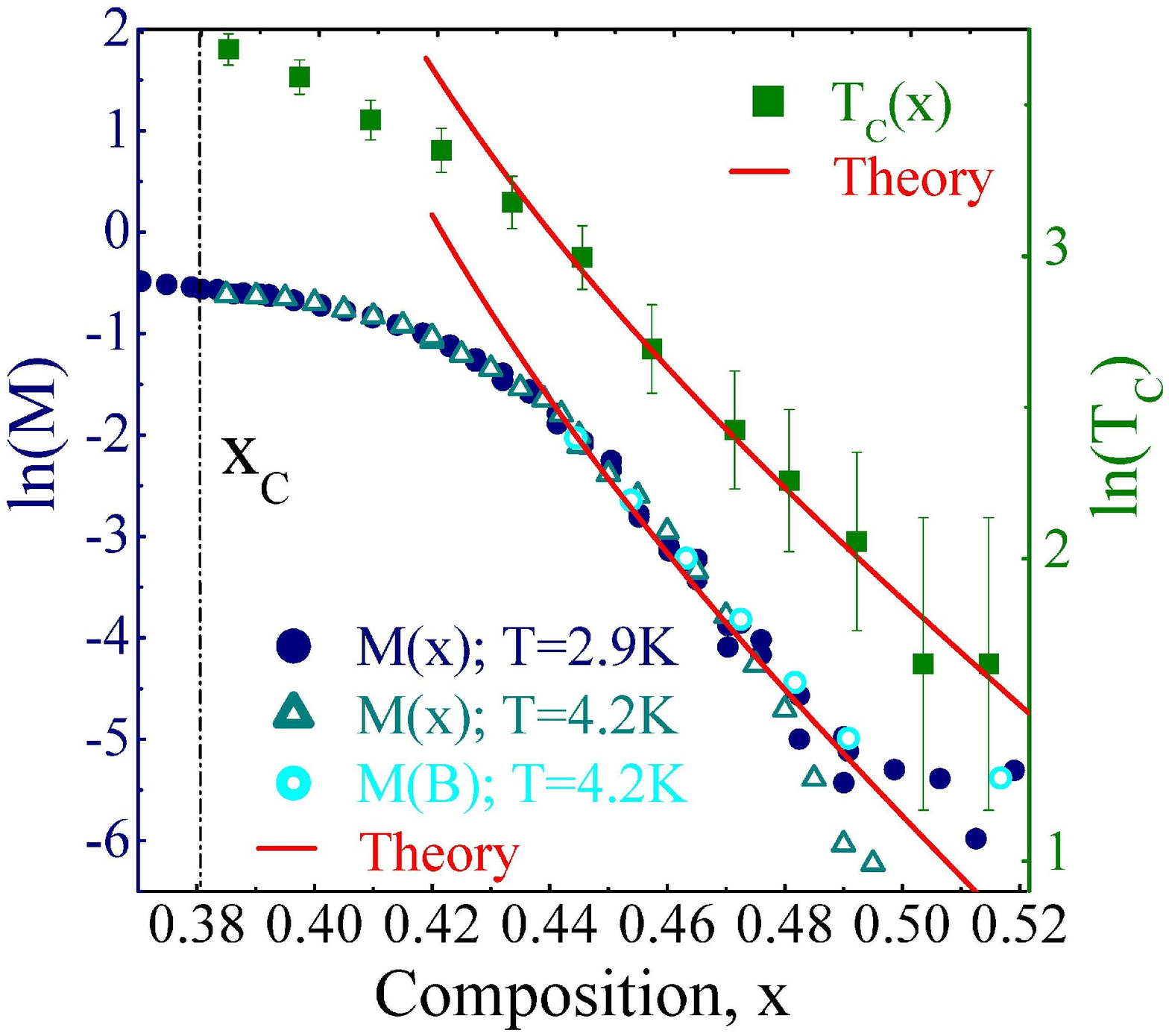}}
\caption{Left: (a) Photographic image of the composition-spread Sr$_{1-x}$Ca$_x$RuO$_3$ epitaxial film.  (b) Contour plot of the magnetization $M$ over the composition-temperature phase diagram. The phase boundary, $T_c(x)$, derived from the susceptibility and magnetization data is shown by the black and white symbols, respectively. (c) Schematic of the magnetism in the tail of the smeared transition. The Sr-rich rare regions form locally ordered clusters whose dynamics freezes, giving rise to an inhomogeneous long-range ferromagnetic order.
Right: Magnetization and transition temperature as functions of the composition in the tail region. The symbols represent the experimental data while solid lines are fits to the theory of Sec.\ \ref{subsec:compo}
(after Ref.\ \cite{Demkoetal12}).
}
\label{fig:SrCaRuO3}
\end{figure}

To study the quantum phase transition in Sr$_{1-x}$Ca$_x$RuO$_3$ with high precision, the group of
K\'ezsm\'arki \cite{Demkoetal12} used a high-quality composition-spread film whose calcium concentration
changes from 13\% to 53\% from one end to the other (see Fig.\ \ref{fig:SrCaRuO3}a).
The magnetic properties were measured using a magneto-optical Kerr microscope which gave a resolution
$\delta x \approx 0.001$ in composition and about $6\times 10^{-3} \mu_B$ per Ru atom in magnetization.

An overview over the results is given in Fig.\ \ref{fig:SrCaRuO3}b. The data show that the critical temperature $T_c(x)$ does not
feature a singular drop at any calcium concentration. Instead, it develops a pronounced tail towards the calcium-rich side of the phase diagram.
A similar tail also appears in the low-temperature magnetization as a function of temperature. These quantities thus provide strong evidence
for the quantum phase transition to be smeared. The right panel of Fig.\ \ref{fig:SrCaRuO3} shows a direct comparison of the magnetization and
critical temperature with the theory of smeared quantum phase transitions discussed in Sec.\ \ref{subsec:compo}. The magnetization data
follow the prediction (\ref{eq:M-exponential}) over 1.5 orders of magnitude in $M$, giving $x_c^0=0.38$. The critical temperature
can be fitted to eq.\ (\ref{eq:Tc-exponential}) using the same $x_c^0$.

Moreover, the smeared transition scenario provides a natural explanation for the strong variations in
the magnetization and phase boundary between the different older experiments. As mentioned at the end
of Sec.\ \ref{subsec:compo} and discussed in detail in
Ref.\ \cite{SNHV12}, smeared phase transitions are particularly sensitive towards spatial disorder correlations. Model
calculations have shown that a disorder correlation length of less than two lattice constants is sufficient
to move the (seeming) critical concentration $x_c$ from 0.5 to 1.

\subsection{Nearly ferromagnetic (Sr$_{1-x}$Ca$_x$)$_3$Ru$_2$O$_7$}

The double-layer ruthenate (Sr$_{1-x}$Ca$_x$)$_3$Ru$_2$O$_7$ has a rich phase diagram
\cite{CMCG97,Quetal08}.
For calcium concentration $x<0.08$, it is in an itinerant metamagnetic state. In the region
$0.08<x<0.4$, the system displays very strong ferromagnetic fluctuations but does not develop
long-range order at low temperatures. Instead, it freezes into a cluster glass state. For $x>0.4$,
the material is an antiferromagnet consisting of ferromagnetic bilayers stacked antiferromagnetically
along the c-axis. The Neel temperature increases from about 30\,K at $x=0.4$ to 56\,K at $x=1$.

Qu et al.\ \cite{Quetal12} performed a detailed analysis of the magnetization $M$ and specific heat $C$ of
(Sr$_{1-x}$Ca$_x$)$_3$Ru$_2$O$_7$ in the composition range $0.08<x<0.4$. Above the cluster glass
freezing temperature (which ranges from less than 1\,K at $x=0.08$ to about 3\,K at $x=0.4$),
both quantities exhibit anomalous power-law singularities. However, while the exponents of $M(H)$ and
$C(T)$ agree, as expected in a quantum Griffiths phase, the exponent characterizing the temperature
dependence $M(T)$ in a small applied field differs significantly. This suggests that the quantum Griffiths
scenario may be insufficient to describe to observed behavior.

The quantum Griffiths scenario assumes that different rare regions are far enough apart to be
effectively independent of each other. The authors of Ref.\ \cite{Quetal12} proposed that the observed
deviations from this scenario may be due to interactions between the rare regions that exist in
itinerant systems \cite{DobrosavljevicMiranda05}. Specifically, the very strong strong ferromagnetic fluctuations
in the composition range $0.08 < x < 0.4$ lead to a ferromagnetic coupling between the rare regions.
The magnetization therefore increases faster upon cooling than the magnetization of isolated
rare regions. Additional theoretical work is needed to provide a quantitative understanding of this
phenomenon.

\subsection{Possible antiferromagnetic Griffiths phase in Ba(Fe$_{1-x}$Mn$_x$)$_2$As$_2$}

As discussed above, a number of explicit observations of rare region (Griffiths) physics in itinerant ferromagnets
have been reported in the last five years. In contrast, analogous observations in itinerant antiferromagnets have been
missing (or, at least, clear-cut verifications of the expected power-law behaviors are not available).
Very recently, Inosov et al.\ \cite{Inosovetal13} reported a possible realization of an antiferromagnetic quantum
Griffiths phase in manganese-doped iron arsenide Ba(Fe$_{1-x}$Mn$_x$)$_2$As$_2$.

The antiferromagnetic spin-density wave order of the parent compound BaFe$_2$As$_2$ is suppressed with increasing
manganese concentration $x$. Above a critical concentration $x_c$ of about 10\%, the character of the magnetic transition
changes: The long-range ordered phase, is preempted by a broad region of phase coexistence between antiferromagnetism,
cluster glass and spin glass order. The authors of Ref.\ \cite{Inosovetal13} interpret the coexistence of magnetically
ordered and paramagnetic clusters as a manifestation of a Griffiths phase and/or the smearing of the antiferromagnetic
quantum critical point due to the random distribution of the manganese atoms.

Further research is required to firmly establish the rare region (Griffiths) scenario in this system. In particular,
various thermodynamic and transport quantities need to be studied and compared in the putative Griffiths region above $x_c$.
Are they controlled by nonuniversal power-laws in $T$ and other control variables? Are their exponents related?
These questions remain a task for the future.

%%%%%%%%%%%%%%%%%%%%%%%%%%%%%%%%%%%%%%%%%%%%%%%%%%%%%%%%%%%%%%%%%%%%%%%%%%%%%%%%
\section{Conclusions}
\label{sec:conclusions}
%%%%%%%%%%%%%%%%%%%%%%%%%%%%%%%%%%%%%%%%%%%%%%%%%%%%%%%%%%%%%%%%%%%%%%%%%%%%%%%%

To summarize, we have reviewed a number of theoretical and experimental developments regarding the unconventional
disorder effects at magnetic quantum phase transitions. The focus has been on novel results that appeared after the
review article \cite{Vojta10}; this means roughly during the last three years.

The theory of smeared quantum phase transitions in itinerant magnets has been extended to the case of transitions
tuned by changing the chemical composition in alloys of type A$_{1-x}$B$_x$. Recent high-precision measurements
demonstrated that the ferromagnetic quantum phase transition in Sr$_{1-x}$Ca$_x$RuO$_3$ is a good example of such a smeared
transition. Advances in theory also showed that quantum Griffiths effects in ferromagnetic metals are
enhanced because the conservation of the ferromagnetic order parameter hampers the relaxation of locally ordered clusters.
The resulting quantum Griffiths singularities are more singular than pure power laws. This may explain the low-temperature
upturns observed in the susceptibility-temperature curves of Ni$_{1-x}$V$_x$. Moreover, a semiclassical theory of various
transport quantities in antiferromagnetic quantum Griffiths phases yielded nonuniversal power-law temperature dependencies
controlled by the same Griffiths exponent that governs the thermodynamics.

On the experimental side, several interesting results have been obtained in addition to the already mentioned smeared
ferromagnetic quantum phase transition in Sr$_{1-x}$Ca$_x$RuO$_3$. The double-layer ruthenate (Sr$_{1-x}$Ca$_x$)$_3$Ru$_2$O$_7$
features a strongly fluctuating nearly ferromagnetic phase for $0.08<x<0.4$ in which magnetization and susceptibility are governed
by nonuniversal power laws. However, the behavior of the exponents does not fully line up with the predictions of the
quantum Griffiths scenario, possibly because of strong interactions between the rare regions. Very recently, a possible
antiferromagnetic Griffiths phase was reported in Ba(Fe$_{1-x}$Mn$_x$)$_2$As$_2$.  This system shows a region of
coexistence between antiferromagnetic and paramagnetic clusters above a manganese concentration of about 10\%.
Further work will be necessary to determine whether this is indeed a manifestation of rare region effects.

In addition to these examples, magnetic cluster effects were also observed in Gd$_5$Ge$_4$ and related intermetallic
compounds \cite{Pereiraetal10,Perezetal11,HadimaniMelnikhovJiles13}. They occur above the sizable critical temperature
of the antiferromagnetic phase transition. The Griffiths singularities should thus be of classical rather than quantum
type. Such classical Griffiths singularities generically require spatial disorder correlations to make them strong enough to be
visible in experiment \cite{Imry77,Vojta06}. The precise nature of these correlations and their effect on observables
requires further experimental and theoretical work. The situation is thus somewhat similar to the Griffiths effects
reported in various manganites,  see the discussion in Ref.\ \cite{Vojta06}.

To conclude, the last three years have seen considerable experimental and theoretical progress in the field of
strongly disordered quantum phase transitions. However, many problems remain at best partially solved. Moreover,
several new questions have emerged that will be attacked in the future.

%%%%%%%%%%%%%%%%%%%%%%%%%%%%%%%%%%%%%%%%%%%%%%%%%%%%%%%%%%%%%%%%%%%%%%%%%%%%%%%%
\ack
This work has been supported in part by the National Science Foundation under Grant No. DMR-1205803.

\section*{References}

\bibliographystyle{iopart-num}
\bibliography{../00Bibtex/rareregions}

\end{document}